\documentstyle[12pt,aasms4]{article}

\lefthead{Torres, Garc\'\i a-Berro \& Isern}
\righthead{Halo white dwarfs}

\begin{document}

\title{Neural Network identification of halo white dwarfs}

\author{Santiago Torres\altaffilmark{1}, 
	Enrique Garc\'\i a-Berro\altaffilmark{2}, and
	Jordi Isern\altaffilmark{3}}

\altaffiltext{1} {Departament de Telecomunicaci\'o i Arquitectura de
	          Computadors, EUP de Matar\'o, Universitat
		  Polit\`ecnica de Catalunya, Avda. Puig Cadafalch 101,
		  08303 Matar\'o, Spain}
\altaffiltext{2} {Departament de F\'{\i}sica Aplicada, Universitat
	          Polit\`ecnica de Catalunya \& Institute for Space 
		  Studies of Catalonia--UPC, Jordi Girona Salgado s/n,
		  M\`odul B-5, Campus Nord, 08034 Barcelona, Spain}
\altaffiltext{3} {Institute for Space Studies of Catalonia--CSIC,
		  Edifici Nexus--104, Gran Capit\`a 2--4, 08034 
	          Barcelona, Spain}

\received{}
\accepted{}

\begin{abstract}
The white dwarf luminosity function has proven to be an excellent 
tool to study some properties of the galactic disk such as its age 
and the past history of the local star formation rate. The existence 
of an observational luminosity function for halo white dwarfs 
could provide valuable information about its age, the time that 
the star formation rate lasted, and could also constrain the
shape of the allowed Initial Mass Functions (IMF). However, the 
main problem is the scarce number of white dwarfs already identified
as halo stars. In this {\sl Letter} we show how an artificial 
intelligence algorithm can be succesfully used to classify the 
population of spectroscopically identified white dwarfs allowing 
us to identify several potential halo white dwarfs and to improve 
the significance of its luminosity function. 
\end{abstract}

\keywords{ stars: white dwarfs -- stars: luminosity function --
Galaxy: stellar content}

\section{Introduction}

Halo white dwarfs have received a continuous interest during almost
one decade from the theoretical (Mochkovitch et al. 1990; Tamanaha 
et al. 1990) and observational points of view. From this last point 
of view Liebert, Dahn \& Monet (1989) studied a high proper motion 
sample and from it derived the first --- and up to now the only one 
available --- halo white dwarf luminosity function. Later Flynn, Gould 
\& Bahcall (1996) and M\'endez et al. (1996) studied the white dwarf 
content of the Hubble Deep Field, after the suggestion of the MACHO 
team that most of the dark matter in the galactic halo could be in 
the form of white dwarfs (Alcock et al. 1997), with negative results. 

The observational white dwarf luminosity function of the halo 
was obtained from a sample of white dwarfs with known parallaxes, 
large proper motions $(2.5^{\prime\prime}\,{\rm yr}^{-1}\ge\mu\ge 
0.8^{\prime\prime}\,{\rm yr}^{-1})$, using a limiting magnitude 
of $m_{\rm V}=19^{\rm mag}$ and assuming that only white dwarfs 
with tangential velocities in excess of 250 km~s$^{-1}$ were 
halo members (Liebert et al. 1989). Consequently only 5 white 
dwarfs contribute to the luminosity function and, thus, the 
statistics is very poor. Besides, in the sample of Liebert et 
al. (1989) there is not any bright halo white dwarf. The absence 
of bright halo white dwarfs in this sample could be due to an 
observational bias but this issue, which bears important 
consequences regarding the IMF of halo stars, remains to 
be studied, since the biased IMFs recently proposed by Adams 
\& Laughlin (1996) and by Chabrier, S\'egretain \& M\'era 
(1996) predict very few bright halo white dwarfs if the halo 
age is large enough (Isern et al. 1998).

In this {\sl Letter} we address the issue of whether or not there 
exist other halo white dwarfs in the existing catalogs and how do 
we identify them. For this purpose we use a neural network technique.
In the end, this will allow us to present a preliminary luminosity 
function of halo white dwarfs and compare it with the theoretical
predictions. 

\section{Method and results}

With the advent of large astronomical databases the need of efficient 
techniques to improve automatic classification strategies has lead to 
a considerable amount of new developments in the field. Among these
techniques the most promising ones are based in artificial intelligence
algorithms. Neural networks have been used successfully in several 
fields such as pattern recognition, financial analysis, biology --- 
see Kohonen (1990) for an excellent review --- and in astronomy. 
For instance, Bezell \& Peng (1998) used these techniques to 
automatically discriminate stars from galaxies, Naim et al. 
(1995) used them to classify galaxies according to their 
morphology, Serra-Ricart et al. (1996) found the fraction 
of binaries in stars clusters, and Hern\'andez-Pajares \& Floris 
(1994) used such techniques to classify populations in the {\sl 
Hipparcos Input Catalogue}.

The common characteristic of all the existing neural network 
classification techniques is the existence of a learning process
very much in the same manner as human experts manually classify.
Generally speaking there are two different approaches: the
supervised and the unsupervised learning methods. The main 
advantage of the last class of methods is that require minimum 
manipulation of the input data and, thus, the results are 
supposedly more reliable. Their leading exponent is the 
so-called Kohonen Self-Organizing Map (SOM). A thourough 
description of this technique is out of the scope of this 
paper. Therefore, we refer the reader to the specific 
literature (Kohonen 1997). However for our purpose it is 
convenient to summarize its basic principle and properties.
The basic principle is to map a multi-dimensional input space 
$(S)$ into a bi-dimensional space $(\Lambda)$. Similar objects 
in $S$ (groups) are mapped in nodes in $\Lambda$. The most 
noticeable property of this procedure is the reduction of 
the dimensionality of the input space allowing, at the same 
time, the identification of groups in the input data and 
the automatic classification of individual objects. Besides, 
neighbor groups in $\Lambda$ have similar properties. 

The catalog of McCook \& Sion (1987) compiles the observational data 
of 1279 white dwarfs. In order to classify the stellar populations 
presumably present in this catalog a set of variables describing their
properties should be adopted. It should be noted that the larger the 
set of variables adopted, the smaller the number of objects that will 
have determinations for all the variables. Conversely, if the number 
of variables in the set is small we could be disregarding valuable 
information. We have adopted a minimal set in order to be able to 
analyse the largest possible number of objects in the catalog. The 
variables adopted in this study are: the absolute visual magnitude 
$M_{\rm V}$, the proper motion $\mu$, the galactic coordinates $(l,b)$, 
the parallax $\pi$, and a color index, $B-V$. This reduces considerably 
the number of objects with {\sl all} the determinations, but allows a 
secure classification. We have found very convenient to use the reduced 
proper motion defined as $H=M_{\rm V}-5\log\pi+5\log\mu$, instead 
of $\mu$ itself because the resulting groups are easier to visualize.

The statistical classification of an observational database usually 
ends up with the detection of groups in the input space that require
an ``a posteriori'' analysis. Since we are interested in detecting 
different stellar populations, simultaneously with the clustering 
process we mix in the input data a synthetic population of tracer 
stars that will allow us to {\sl label} the groups detected by the 
classification algorithm as halo, disk or intermediate population. 
The results of the classification procedure are not sensitive to 
the fine details of these synthetic populations, except to the IMF. 
These synthetic tracer stars have been produced using a Monte Carlo 
(MC) simulator. The description of the MC simulation of the disk 
population can be found in Garc\'\i a-Berro \& Torres (1997) or 
Garc\'\i a-Berro et al. (1998). A comprehensive discussion of the 
results of the MC simulation of the halo population will be published 
elsewhere. However, and for the sake of completeness, a brief 
summary of the inputs is given here. We have adopted a standard, 
Salpeter-like, IMF (Salpeter 1961). The halo was supposed to be 
formed 14 Gyr ago in an intense burst of star formation of 1 Gyr 
of duration. The stars are randomly distributed in a sphere of 
radius 200 pc centered in the sun according to a density profile 
given by the expression $\rho(r)\propto(a^2+R^2_{\sun})/(a^2+r^2)$, 
being $r$ the galactocentric radius, $a\approx 5$ kpc and $R_{\sun}
=8.5$ kpc. The velocities of the tracer stars where randomly drawn 
according to normal distributions for both the radial and the 
tangential components, with velocity dispersions as given in 
Markovi\'c \& Sommer-Larsen (1997); the adopted rotation velocity 
$V_{\rm c}$ is 220 km~s$^{-1}$. The remaining inputs were the same 
adopted in Garc\'\i a-Berro et al. (1998). In order to reproduce 
accurately the properties of the real catalog both MC simulations 
were required to meet the additional set of criteria: $\delta\ge 
0^\circ$, $8.5\le M_{\rm V}\le 16.5$, $\mu\le 4.1^{\prime\prime}
\,{\rm yr}^{-1}$ and $0.006^{\prime\prime}\le\pi\le 0.376^{\prime
\prime}$, which are derived from the subset of 232 white dwarfs 
which have all the determinations. An added value of the above 
described procedure of mixing tracer and real stars is that in 
this way we can check the accuracy of the classification algorithm 
and the quality of the MC simulations. 

The simulated samples mimic fairly well the observational sample
as can be seen in figure 1, where the results of the MC simulations 
for the disk and the halo are compared with the observational sample 
in the reduced proper motion-color diagram. As can be seen in this 
diagram the two simulated samples are easily visualized. Similar 
diagrams can be produced for each pair of variables and the results 
of the MC simulations compare equally well with the real data. 

We have run the public domain neural network software SOM\_PAK 
(available at http://www.cis.hut.fi/nnrc/som\_pak/) with a 
catalog constructed as described above. The SOM of the input 
catalog, after three passes over the entire sample and with 
a grid of $5\times 5$ nodes is shown in figure 2. The groups 
have been assigned either to halo (``H'') or disk (``D'') 
populations if the percentage of tracer stars of one of the 
populations was larger than 70\%. In the groups labeled as ``I'' 
(intermediate population) neither the halo nor the disk tracers 
were in excess of this recognition percentage. As can be seen in 
figure 2 all the halo groups are close neighbors and, furthermore, 
the intermediate population groups are surrounded by halo and disk 
groups. A good measure of the overall quality of the classification 
scheme can be obtained by checking how many of the synthetic stars
are missclassified. This results in the following confusion matrix 

$$
C=
\left(
\begin{array}{cc}
0.98 &0.03\\
0.02 &0.97\\
\end{array}
\right)
$$

\noindent
where the matrix element $C_{11}$ indicates the percentage of disk
tracers classified in disk groups, $C_{21}$ is the percentage of disk
tracers missclassified in halo groups, and so on. This matrix is 
very close to unity, and thus the classification seems to be secure. 
More confidence in this classification comes from the fact that the
vast majority of old disk white dwarfs in the sample of Liebert, Dahn 
\& Monet (1988) are in the groups (0,2) and (2,1) which are labeled 
as intermediate population. Moreover, LHS~56, LHS~147 and LHS~291 
belong to the group (1,0) which clearly is a halo group, and LHS~2984 
belongs to the group (0,0), all these objects were already identified 
as halo members by Liebert et al. (1989), and were used to build 
their halo white dwarf luminosity function. The only object of the 
sample of Liebert et al. (1989) missclassified is LHS~282 which is 
classified in the group (0,2), which is intermediate population. All 
this evidence points in the same direction: the classification is 
correct. The percentages of halo tracers in the groups labeled as halo 
can be found for each of the halo groups in figure 2. Since all of 
them are larger than 80\% all these groups can be securely labeled 
as halo. However, and for the sake of reliability we have only 
identified as halo candidates those white dwarfs belonging to 
groups which do not have a disk neighbor, namely (0,0), (0,1), 
(1,0) and (2,0). One interesting property of these white dwarfs 
is that all of them have $M_{\rm V}\le 14$ and only 4 have proper 
motions in excess of $1.0^{\prime\prime}\,{\rm yr}^{-1}$, being 
the average $\langle\mu\rangle=0.87^{\prime\prime}\,{\rm yr}^{-1}$. 
However, most of them have $\pi\le 0.03^{\prime\prime}$, and are 
clustered around $\pi\sim 0.01^{\prime\prime}$, leading to 
tangential velocities in excess of 200 km~s$^{-1}$ for 11 of our 
candidates. Only one candidate has a tangential velocity smaller 
than 100 km~s$^{-1}$. Therefore the detected population is 
intrinsically bright and distant. The halo white dwarf candidates 
detected here can be found in table 1. 

\placetable{tab1}

We have used the $1/V_{\rm max}$ method to derive a luminosity 
function of halo white dwarfs with the candidates found so far. 
The adopted criteria for deriving such a luminosity function
were those of Oswalt et al. (1996). The result is shown in figure 
3 as solid triangles. The error bars have been computed as in 
Liebert et al. (1988). The number of objects in each luminosity
bin is shown on top of the corresponding error bar. The value of 
$\langle V/V_{\rm max}\rangle=0.115$ indicates that the sample is 
not complete.  Thus, this halo white dwarf luminosity function should 
be considered as preliminary although represents a considerable 
improvement over that of Liebert et al. (1989). According to the 
previous discussion, the most obvious feature of this luminosity 
function is the detection of halo candidates for the brightest 
luminosity bins. The faintest halo candidate found by Liebert et 
al. (1989), LHS~282, has been classified as intermediate population 
and, therefore, the faintest bin in their luminosity function is 
absent in ours. Some other small differences for the faintest bins 
are due to the binning procedure and to the bolometric magnitudes 
assigned to individual objects --- we have used the bolometric 
corrections of Bergeron et al. (1995), whereas Liebert et al. 
(1989) used blackbody bolometric corrections. 

For comparison purposes in figure 3 we also show two theoretical 
luminosity functions computed with a standard IMF. The adopted ages 
were in both cases 14 Gyr, and the durations of the bursts were 1 
Gyr. The solid line corresponds to a luminosity function computed 
with a H envelope (Wood 1995) whereas the dotted line shows the 
luminosity function obtained using a He dominated envelope (Wood 
\& Winget 1989). The method used to compute both luminosity functions 
is that of Isern et al. (1998), where the rest of the details of the 
adopted inputs can be found. Also shown in figure 3 is the detection 
limit of very faint white dwarfs (open triangles) of Liebert et al. 
(1988). Both luminosity functions have been normalized to the bin 
with the smallest error bars and present a reasonable agreement with 
both sets of observational data. 

\section{Conclusions}

We have shown that an artificial intelligence algorithm is able
to classify the catalog of spectroscopically identified white 
dwarfs and ultimately detect several potential halo white dwarfs. 
Some of these white dwarfs were already proposed as halo objects 
by Liebert et al. (1989). We have found as well that our halo 
candidates are bright and distant, and that most of them have 
large tangential velocities. Using the $1/V_{\rm max}$ method 
we have computed a preliminary luminosity function of halo 
white dwarfs. We have found a value of $\langle V/V_{\rm max}\rangle
=0.115$ which indicates that our luminosity function is still 
uncomplete. However, this luminosity function largely improves 
the previous one by Liebert et al. (1989). We have also compared 
this luminosity function with the theoretical predictions, and we 
have found a fair agreement with the luminosity functions computed 
with a standard IMF. 

\vspace{1cm}

\noindent
{\em Acknowledgements} This work has been supported by DGICYT grants 
		       PB94-0111, PB94--0827-C02-02, and by the CIRIT 
		       grant GRC94-8001.

\newpage

\begin{deluxetable}{lccccrc}
\tablewidth{0 pt}
\tablecaption{Halo white dwarf candidates identified using the neural 
network algorithm, along with their corresponding group and properties.
The stars already identified in Liebert et al. (1989) are marked with
and asterisk.
\label{tab1}}
\tablehead{
\colhead{Name} &
\colhead{Group} &
\colhead{$M_{\rm V}$} &
\colhead{$\mu\,(^{\prime\prime}\,{\rm yr}^{-1})$} &
\colhead{$\pi\,(^{\prime\prime})$} &
\colhead{$B-V$} &
\colhead{Sp. Type}
}
\tablewidth{0 pt}
\startdata
LHS~2984$^*$   & (0,0) & 11.62 & 0.930 & 0.015 & $ 0.03$ & DA \\
LHS~3007       & (0,0) & 13.06 & 0.636 & 0.028 & $ 0.29$ & DA \\
G~028-027      & (0,0) & 12.41 & 0.281 & 0.003 & $ 0.03$ & DQ \\
G~098-018      & (0,0) & 11.81 & 0.426 & 0.003 & $ 0.38$ & DA \\
G~138-056      & (0,0) & 13.34 & 0.692 & 0.006 & $ 0.37$ & DA \\
G~184-012      & (0,0) & 13.18 & 0.427 & 0.017 & $ 0.26$ & DC \\
LP~640-069     & (0,0) & 12.75 & 0.284 & 0.009 & $ 0.29$ & DA \\
LHS~56$^*$     & (0,1) & 13.51 & 3.599 & 0.069 & $ 0.36$ & DA \\
LHS~147$^*$    & (0,1) & 13.64 & 2.474 & 0.016 & $ 0.40$ & DC \\
LHS~151        & (0,1) & 13.46 & 1.142 & 0.053 & $ 0.33$ & DA \\
LHS~291$^*$    & (0,1) & 13.39 & 1.765 & 0.012 & $ 0.11$ & DQ \\
LHS~529        & (0,1) & 13.94 & 1.281 & 0.046 & $ 0.64$ & DA \\ 
LHS~1927       & (1,0) & 11.41 & 0.661 & 0.009 & $ 0.11$ & DA \\
G~038-004      & (1,0) & 12.31 & 0.428 & 0.010 & $ 0.17$ & DA \\
LHS~3146       & (2,0) & 11.88 & 0.579 & 0.024 & $ 0.17$ & DA \\
G~021-015      & (2,0) & 11.54 & 0.390 & 0.015 & $ 0.05$ & DA \\
G~035-026      & (2,0) & 11.12 & 0.335 & 0.007 & $-0.14$ & DA \\
G~128-072      & (2,0) & 12.53 & 0.457 & 0.025 & $ 0.21$ & DA \\
G~271-106      & (2,0) & 11.77 & 0.396 & 0.014 & $ 0.18$ & DA \\
GR~363         & (2,0) & 11.39 & 0.133 & 0.003 & $-0.03$ & DA \\
\enddata
\nl
\end{deluxetable}

\newpage

\begin{figure}
\vspace{11cm}
\includegraphics{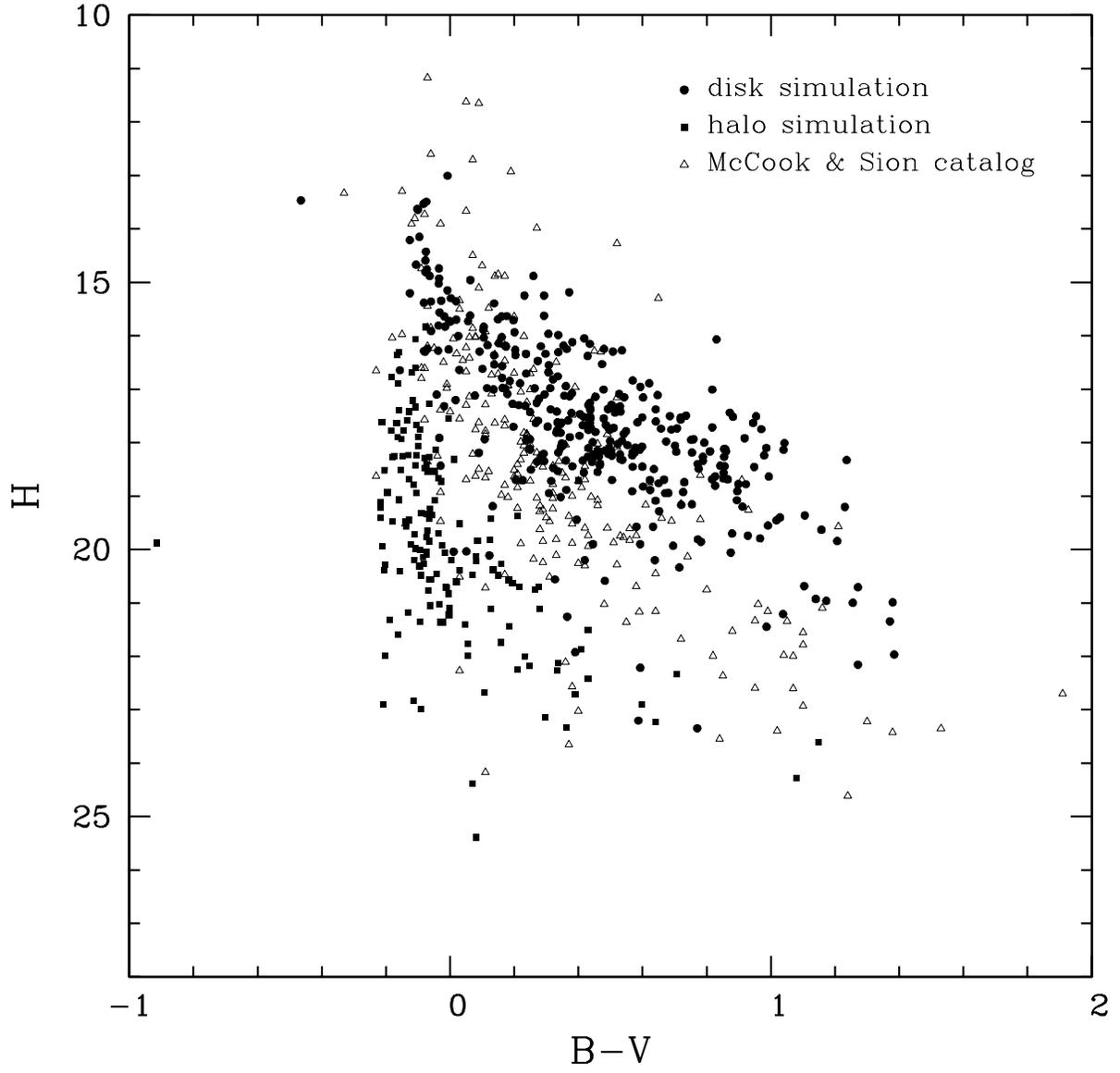}
\caption{Reduced proper motion-color diagram for the MC simulations
of the disk (solid dots) and the halo (solid squares) and of the 
observational sub-sample (open triangles).}
\end{figure}

\begin{figure}
\vspace{11cm}
\includegraphics{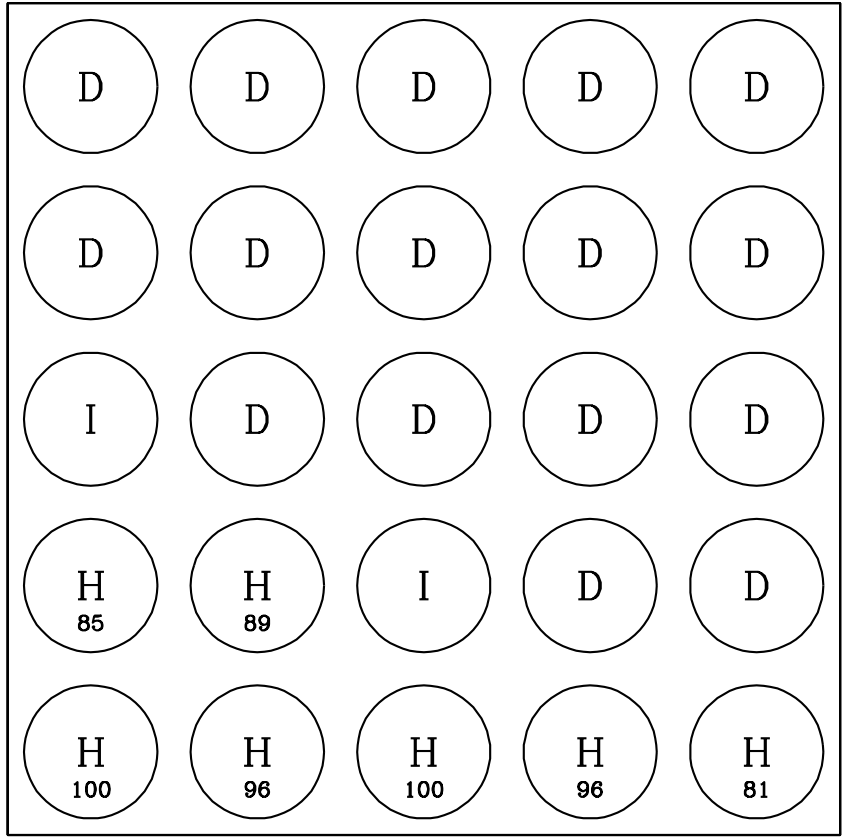}
\caption{Self-organizing map of the sample of white dwarfs, see
text for details. The group (0,0) is located in the lower left corner
of the diagram and the group (4,4) is located in the upper right
corner. As a rule of thumb $H$ increases from right to left and
$M_{\rm V}$ decreases downwards in the diagram.\label{fig2}}
\end{figure}

\begin{figure}
\vspace{11cm}
\includegraphics{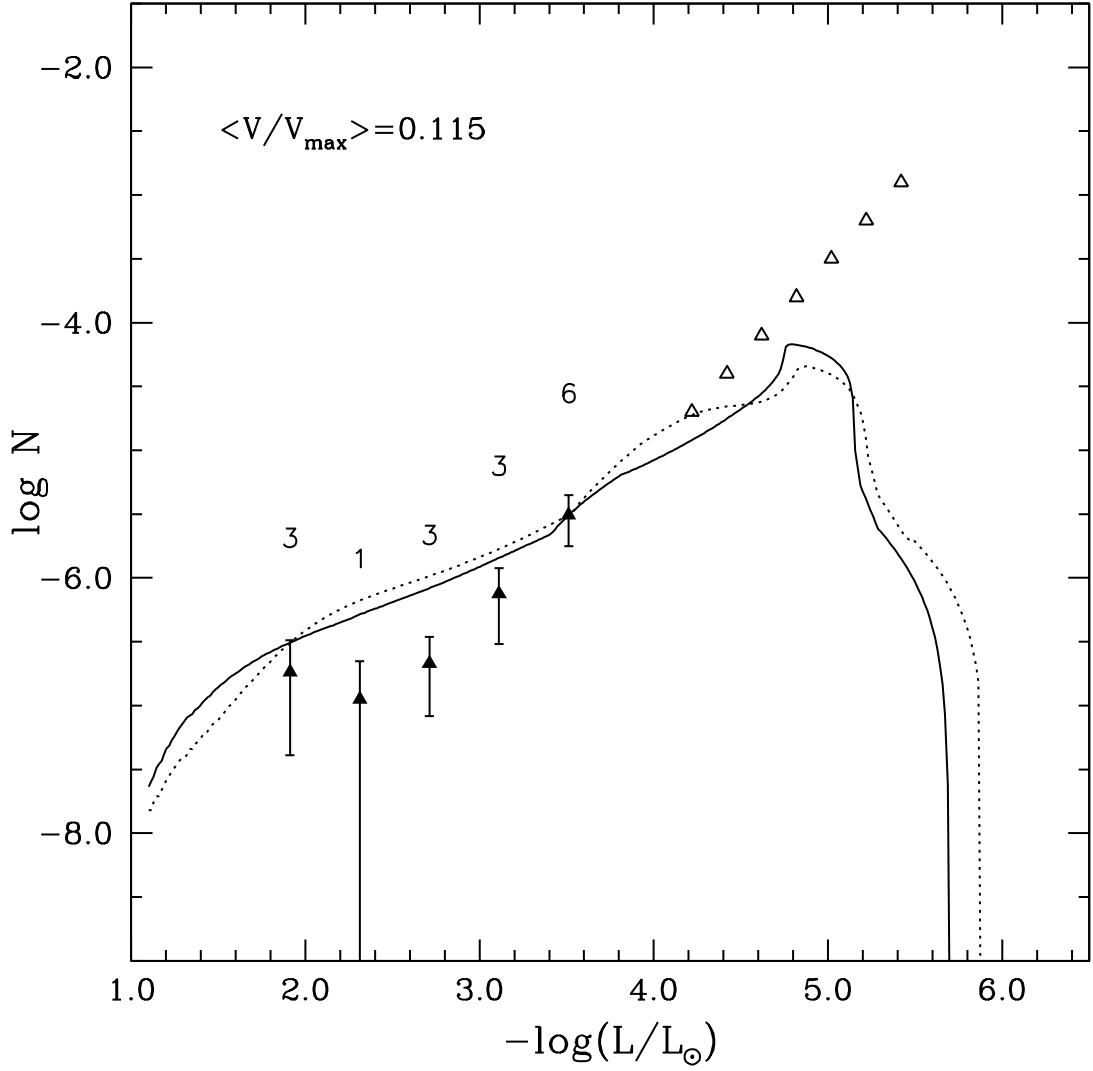}
\caption{White dwarf luminosity function obtained using the 
$1/V_{\rm max}$ method with the halo white dwarf candidates found 
in this work --- solid triangles --- and the detection limit of 
Liebert et al.  (1988) --- open triangles. The solid and the 
dotted lines are theoretical white dwarf luminosity functions
obtained assuming a standard IMF and thick H and He dominated
envelopes, respectively.\label{fig3}}
\end{figure}

\end{document}